\documentclass[12pt,preprint]{aastex}

\def\Msun{\hbox{\it M$_\odot$}}

\def\Gyr{\hbox{\it Gyr}}
\def\pc{\hbox{\it pc}}

\def\kms{\hbox{\it km$\,$s$^{-1}$}}

\def\simgr{\mathrel{\hbox{\rlap{\hbox{\lower4pt\hbox{$\sim$}}}\hbox{$>$}}}}
\makeatletter
\def\jnl@aj{AJ}
\ifx\revtex@jnl\jnl@aj\let\tablebreak=\\\fi
\makeatother
\received{}
\revised{REVISION DATE}
\accepted{}
\journalid{VOL}{JOURNAL DATE}
\articleid{START PAGE}{END PAGE}
\paperid{MANUSCRIPT ID}
\cpright{TYPE}{YEAR}\ccc{CODE}

\begin{document}

\title{High Precision Stellar Radial Velocities in the Galactic Center
\footnote{Data presented herein were obtained at the W.M. Keck
Observatory,  which is operated as a scientific partnership among the
California Institute of Technology,  the University of California and
the National Aeronautics and Space  Administration.  The Observatory
was made possible by the generous financial support of the W.M. Keck
Foundation.}}

\author{
Donald F. Figer\altaffilmark{2,3}, Diane Gilmore\altaffilmark{2}, Sungsoo S. Kim\altaffilmark{4},  \\
Mark Morris\altaffilmark{5}, E. E. Becklin\altaffilmark{5}, Ian S. McLean\altaffilmark{5}, \\
Andrea M. Gilbert\altaffilmark{6}, James R. Graham\altaffilmark{6}, \\
James E. Larkin\altaffilmark{5}, N. A. Levenson\altaffilmark{7},
Harry I. Teplitz\altaffilmark{8}}

\email{figer@stsci.edu}

\altaffiltext{2}{Space Telescope Science Institute,
	3700 San Martin Drive, Baltimore, MD 21218; figer@stsci.edu }
\altaffiltext{3}{Department of Physics and Astronomy, Johns Hopkins University,
	Baltimore, MD  21218}
\altaffiltext{4}{Kyung Hee University, Dept. of Astronomy \& Space Science, Yongin-shi, Kyungki-do 449-701, Korea}
\altaffiltext{5}{Department of Physics and Astronomy, University of California, Los Angeles,
	Division of Astronomy, Los Angeles, CA, 90095-1562 }
\altaffiltext{6}{Department of Astronomy, University of California, Berkeley, 601 Campbell Hall,
	Berkeley, CA, 94720-3411}
\altaffiltext{7}{University of Kentucky, Department of Physics and Astronomy, Lexington, KY 40506-0055, USA}
\altaffiltext{8}{SIRTF Science Center, California Institute of Technology, 220-6, Pasadena, CA, 91125}

\begin{abstract}
We present radial velocities for 85 cool stars projected
onto the central parsec of the Galaxy. The majority of these velocities have relative errors of $\sim$1~\kms, or
a factor of $\sim$30-100 smaller than those previously obtained with proper
motion or other radial velocity measurements for a similar stellar sample.
The error in a typical individual stellar velocity, including all sources of uncertainty, is 1.7~\kms.
Two similar data sets were obtained one month apart, and the total error in the relative velocities is 0.80~\kms\
in the case where an object is common to both data sets.
The data are used to characterize the velocity distribution of
the old population in the Galctic Center. We find that
the stars have a Gaussian velocity distribution with a mean heliocentric velocity of $-$10.1$\pm$11.0~\kms\
(blueshifted) and a standard deviation of 100.9$\pm7.7$~\kms; the mean velocity
of the sample is consistent with no bulk line-of-sight motion with respect to
the Local Standard of Rest. At the 1 sigma level, the
data are consistent with a symmetric velocity distribution about any arbitrary axis
in the plane of the sky. We find evidence for a flattening in the distribution of late-type stars within a
radius of $\sim$0.4~\pc, and infer a volume density distribution of r$^{-1/4}$ in this region.
Finally, we establish a first epoch of radial
velocity measurements which can be compared to subsequent epochs to measure
small accelerations (1~\kms~yr$^{-1}$),
corresponding to the magnitude expected over a timespan of several years for stars nearest to Sgr~A*.
\end{abstract}


\section{Introduction}

Evidence supporting the presence of a supermassive black hole in our Galactic Center (GC) is now
very strong \citep{ghe00,sch02}, and is certainly the strongest for any such object.
While the current estimates of the central dark mass are nearly identical to
the original estimates, M $\approx$
3(10$^6$)\Msun\ \citep{lac82,ser85}, estimates of the bounding volume containing this mass have grown much
smaller over the years. The resultant density estimates have grown from 10$^6$~\Msun\pc$^{-3}$ to $>$10$^{17}$~\Msun\pc$^{-3}$,
suggesting that the central dark mass is indeed a black hole.
All of the mass determinations rely upon gas or stellar velocity measurements, either along
the line of sight or in the plane of the sky, and are valid in the case that the motions are dominated
by gravitational forces.

The ionized gas in the central parsec orbits the black hole, and its bulk motion can be described
by a model that assumes a large central mass \citep{ser85,gus87,ser88,lac91,rob93,her93}.
The stellar motions in the central parsec
exhibit ordered rotation for the young population \citep{gen00,pau01,gen03} and an isotropic distribution
for the old population \citep{gen00}. \citet{mcg89} and \citet{sel90} use the line-of-sight motions of the old
late-type stars to infer presence of the central dark mass; simliar experiments were done
by \citet{hal96a} and \citet{sah96}. Finally, two groups have independently developed
proper motion analyses that provide the best evidence
supporting a super-massive black hole in the GC \citep{gen96,gen97,ghe98,gen00,ghe00,eck02,sch02,ghe03,gen03}.

Supermassive black holes have been inferred to exist at the dynamical centers of a number of
galaxies having bulges, or bulge-like populations, and their presence may be ubiquitous in such
galaxies \citep{fer00,tre02}.
It is important to establish the existence, and measure the properties, of supermassive black holes, given their
potentially prominent role in controlling the energy output of galaxies and their
diagnostic value in probing galactic evolution. In addition, it is also important to
determine whether objects other than a supermassive black hole might contribute to
enclosed mass estimates, i.e.\ various forms of dark matter.


In this paper, we report the first data set obtained from a new observational program
to determine the mass distribution
in the central parsec, as determined from line-of-sight velocities of late-type stars.
In this program, we obtained high spectral resolution data of 85 cool stars
throughout the central parsec using NIRSPEC on Keck II. We intend to combine these and similar data
obtained in the future in order to measure accelerations, thereby constraining the mass distribution
in the central parsec.

\section{Observations and Data Reduction}

\subsection{Observations}

The observations were obtained on June 4, 1999, and July 4, 1999 with NIRSPEC, the
facility near-infrared spectrometer, on the Keck II telescope \citep{mcl98}, in high
resolution mode covering {\it K}-band wavelengths ($1.98~\micron$ to
$2.32~\micron$). The long slit (24\arcsec) was used in a north-south
orientation, and the telescope was offset by a fraction of a slit width
to the west between exposures. Two datacubes were built from the
spectra obtained on the two nights. The field observed in July 
is almost entirely contained within the
field imaged in June. Spectra of 40 of the same stars, 22 of them cool, appear
in both data cubes. A log of observations is given in Table~1. The slit viewing
camera (SCAM) was used to obtain images simultaneously with the
spectra. These images make it easy to determine the slit orientation
on the sky when the spectra were obtained; Figures~\ref{fig-gc1scam_slits}a
and \ref{fig-gc1scam_slits}b show the inferred slit
positions during all spectral exposures. From the SCAM images, we
estimate seeing (FWHM) of 0$\farcs$6 on the first night and 0$\farcs$4 on the
second night. The plate scales for spectrometer and imager
(0$\farcs$18) were taken from \citet{fig00}. We
chose to use the 5-pixel-wide slit (0\farcs72) for the first slit scan
and 3-pixel-wide slit (0\farcs43) for the second slit scan in order to
match the seeing on the respective nights. The corresponding resolving
power was R$\sim$14,000 in June, and R$\sim$23,300 in July, measured
from sky OH and arc lamp lines of known wavelengths in the two datasets. The slit positions were
nearly parallel in the June scan, aligned with the long axis along the north-south direction, while they were quite skewed in the latter
part of the July scan because of problems with telescope tracking and/or image
rotator control during these commissioning runs.

The NIRSPEC cross-disperser and the NIRSPEC-7 and NIRSPEC-6 filters were used to
image six echelle orders onto the 1024$^2$-pixel InSb detector
field of view. The approximate spectral ranges covered in these orders
are listed in Table~2. Note that the two filters were used in concert on the second night in order
to ensure complete rejection of light in unwanted cross-disperser orders.
Some prominent features in our data set include diffuse and stellar lines of:
Br-$\gamma$ emission at 2.1661$~\micron$ in
order 35, \ion{Fe}{3} at 2.2178$~\micron$ in order 34 of the June data, \ion{He}{1}
at 2.0581$~\micron$ in order 37, and the edge of the CO(2-1)
transition at 2.2935$~\micron$ in order 33.

Quintuplet Star \#3  (hereafter ``Q3''), which is featureless in this
spectral region \citep{fig98}, was observed as a telluric
standard \citep{mon94}. Arc lamps containing Ar, Ne, Kr, and
Xe, and spectra of the sky, were observed to set the wavelength scale. In addition, a continuum
lamp was observed through an etalon filter in order to produce an
accurate wavelength scale inbetween arc lamp and sky lines
(predominantly from OH). A field relatively devoid of stars (``dark spot'')
(RA~17$^{\rm h}$~44$^{\rm m}$~49$\fs$8,
DEC~$-$28$^{\arcdeg}$~54$^{\arcmin}$~6$\farcs$8~, J2000) was observed
to provide a dark current plus bias plus background image. A
quartz-tungsten-halogen (QTH) lamp was observed to provide a ``flat''
image.

\subsection{Data Reduction}

We removed sky emission, dark current, and residual bias, by subtracting
images of the dark spot.
Due to the changing sky levels during the scan, the sky in the dark spot was scaled
before subtracting it from target images.  The scale factor was
chosen such that it produced the minimum variance of
residual sky line features in selected regions of the image
after flat-fielding. The images were then ``flattened'' by dividing
them by the QTH lamp image. These
flattened images were then cleaned of bad pixels using a two-pass
procedure. First, we replaced any pixel with a value 5$\sigma$ above
neighbors within a 5 by 5 pixel area by the median of its neighbors in
that area. Second, we replaced pixels with values that
were higher or lower than the values of both immediate neighbors in
the dispersion direction, and deviated
by more than 10 times the poisson noise from those two neighbors. The results of these
procedures are shown in Figure~\ref{fig-04jus0119_red.fits}. The image was obtained
with the slit positioned approximately north-south and centered on SgrA*.
The echelle order number
increases toward the bottom of the image, and north on the sky is downward in the slit.
One can clearly see the diffuse Br-$\gamma$ near the center of the image, and even the
ionized ``tail'' associated with IRS7 \citep{yus91}.

Several images of the telluric standard, Q3, were taken in the
same setup that was used to obtain the target images. In each of the images, spectra were
imaged onto different rows of the detector. We then removed dark current,
sky emission, and residual bias, by subtracting one image from the
other. These images were further reduced by
flat-fielding and cleaning deviant pixels, as described above for the
target images.

The etalon, arc and sky images were also reduced as described above.
The reduced versions of these images were used for preliminary rectification
of the warped echelle orders.

\subsection{Source Identification}
Figure~\ref{fig-scam_ids} shows a SCAM image of the slit-scan region overplotted with
identifiers taken from Table~4.
Sources in the sample were identified based on their distance from the closest, brightest source
in \citet{gen00}. Note that the sample in \citet{gen00} is drawn from the sample
in \citet{gen96}.
In cases where no obvious counterpart in \citet{gen00} could be
found, we label the star according to the ID number in Table~4. All 85 cool stars
in the table are labelled in the figure, and the spectroscopic sample is $\approx$50\% complete
to K=13.

\subsection{Spectral Image Rectification}
We rectified the data format into orthogonal axes of wavelength and
(approximately) declination by mapping a set of observed features in the
warped images onto a set of dewarped grid points. Typically, 15 to 20 spectral
lines of known wavelength were used in each order as wavelength
fiducials; the zero-point wavelength was set by sky OH lines in the target
spectra.
Two stellar continuum spectra and two slit edges traced across the length of
the dispersion direction served as spatial references. Figure~\ref{fig-gc119_rect.fits} shows
the result of this process, which we explain in more detail below.

In general, we traced arc, sky, and etalon lines as follows. Each line was
divided into 10 to 20 equally sized samples along the slit length, each sample
containing 10 rows of data. Rows within each sample were averaged together, and
the location of the centroid along the x-axis was recorded. We then
used a 3rd order polynomial to fit the location of the centroids as a function
of location along the y axis. In cases where the
centroiding was less robust, due to a nearby unidentified spectral line,
or when remaining bad pixels affected the centroiding, the order of the polynomial fit was
reduced and/or points near the edges were deleted.
We traced spatial locations along the slit length as a function of
location along the wavelength axis in much the same way, except that the
centroids were computed along the y axis.

Given the spatial and spectral centroids from the preceeding process, we
then rectified the data in three stages: 1) rectify images
using the arc and sky lines, 2) measure the etalon line wavelengths in the
rectified image, and obtain a solution to the etalon equation, and 3) use
the analytically-determined wavelengths of the etalon lines to produce a better
rectification matrix.

In the first stage, we produced a solution using an order containing many (15 to 20) arc
and sky lines. Order 35 of the June dataset and order 36 of the
July dataset were selected. OH line wavelengths were taken from
\citet{abr94}. The arc line wavelengths were obtained from the
National Institute of Standards and Technology (NIST) Atomic Spectra
Database\footnote{http://physics.nist.gov/PhysRefData/ASD1/nist-atomic-spectra.html}.
The arrays of coefficients from spatial and spectral fits were then
used to produce a mapping between points in
the warped frame, and points in the dewarped frame. We then fit a two-dimensional
polynomial, of second or third order, to these points, resulting
in a list of transformation coefficients. Finally, we rectified
the selected orders of the etalon image.

In the second stage, we measured
the wavelengths of the etalon lines using ``SPLOT'' in IRAF\footnote {IRAF
is distributed by the National Optical Astronomy Observatories, which
are operated by the Association of Universities for Research in
Astronomy, Inc., under cooperative agreement with the National Science
Foundation.}. This provided preliminary wavelengths for each
etalon line. Exact etalon wavelengths are given by the etalon equation,
i.e.\ $\lambda$=t/2n, where t is the thickness of the air gap of the etalon
and n is the etalon order number. Each etalon line will have its own equation
that relates its specific wavelength to its specific order number.
We simultaneously solved a set of etalon equations for the etalon
lines we fit using the wavelength solution from stage one, under the constraint that all equations must
use the same thickness, and that the order numbers be increasing integers
from longer to shorter wavelengths. We varied the thickness and order numbers
until the wavelengths predicted by our fit were closest to the wavelengths
found in stage one.

In stage three, we used our etalon wavelengths from stage two to fit the
etalon lines in all echelle orders. Together with the same spatial
information used to produce the first stage rectification matrix, a
new rectification matrix was produced. The improved quality of the etalon
lines allowed for a higher order (fourth or fifth order) polynomial fit
to the etalon mapping between warped and dewarped points. The new
matricies were applied to the appropriate spectral orders of target
images to rectify them.

\subsection{Building the Data Cubes}
We extracted stellar spectra, typically four to eight per slit pointing, using ``APALL'' in IRAF.
Spectra of the same object observed in multiple telescope pointings were coadded,
after first shifting each spectrum to account for slight variations in the object's position along
the slit width (dispersion) direction. This was done by first translating each spectrum along the wavelength direction
by an amount equal to the offset between it and the spectrum of IRS7.
To obtain this offset, we cross-correlated the spectra of the target and IRS7 near 2.05~\micron,
where many telluric features are present. We applied
another shift to account for the fact that IRS7 was not likely in the center
of the slit when it was observed. To estimate this shift, we measured the
median of the offsets of all target stars with respect to the spectrum of IRS7, and
assumed that the distribution of positions within the slit was random, i.e.\
we added the negative of this offset to each target spectrum. The total shift to
compensate for target position within the slit along the dispersion direction was
typically a few tenths of a slit width, or equivalently, approximately one pixel.

In using the procedure above, we are relying on the fact that wavelengths of the telluric features do not
change. In addition, we rely on the fact that the grating remained in the same position
throughout the observing sequence, and that variable flexure in the instrument was
inconsequential; note that NIRSPEC sits on the Nasmyth deck and therefore has a
constant orientation with respect to gravity. Our data do not suffer from the problem
that currently plagues some NIRSPEC echelle data, that of a wavelength shift induced by grating
motion when the image rotator is slewed.

Spectra of the source Q3 were extracted and shifted to the center of the slit,
median-combined, and then normalized to 1 in order 36.
Target spectra were then divided by these final
standard spectra to remove telluric lines and to normalize and remove
the shape of the stellar continuum, making it easier to compare the CO
bandhead and other features of interest.

On the first night of observations, we obtained spectra for 152 stars, including 102 cool stars,
34 hot stars, and 16 stars that have featureless or indeterminate spectra. We have separated the cool stars
into two groups, based upon quality of the data, i.e.\ 78 are of high quality and 24
are of low quality. On the second night of observations, we obtained spectra for
68 stars, including 38 cool stars, 16 hot stars, and 14 stars that have featureless
or indeterminate spectra. Of the cool stars, 29 have high quality spectra, and 9 have
low quality spectra.

In this paper, we present the velocities of only
the cool stars, given that their spectra can be analyzed to produce the most precise velocities;
in addition, the cool star orbits are much more likely to be thermalized and thus better
suited for use in probing the local matter distribution than the hot star orbits. We only
include the stars with high quality spectra.
Unfortunately, the hot star spectra usually have intrinsic features, produced in photospheres or winds, that are confused
with features from nearby ionized gas, i.e.\ gas in the GC mini-spiral. In addition,
their features are sometimes broad and/or asymmetric, both of which lead to
much larger errors in determining their velocities compared to those obtainable
from cool star spectra. We will report on our efforts to extract precise
velocities from the hot star spectra in a separate paper.

\subsection{Cool star velocities}
Most ($\sim$60\%) stars in our sample are
cool, showing varying amounts of absorption due to CO, as well as many other
atomic and molecular species. We used the CO bandhead (edge rest wavelength is 2.2935$~\micron$) to
determine velocities by cross-correlating spectra of the sample stars with vacuum rest frame spectrum
of Arcturus \citep{hin95} (R=100,000), correcting for
Earth motion toward the Galactic Center and differences in spectral type between the
sample stars and Arcturus.

We corrected Arcturus' spectrum to the rest frame by maximizing the cross-correlated
power between it and a spectrum of the Sun in the wavenumber region from 4300~cm$^{-1}$ to 4375~cm$^{-1}$,
after first resampling the solar spectrum \citep{liv91} so that both spectra had the same effective resolution (R$\sim$100,000).
We believe that the adjustment to the rest frame is accurate to within one tenth of this resolution, or $\pm$0.30~\kms.
The rest spectrum of Arcturus was then resampled and smoothed to match the spacing and resolution of the GC target spectrum.
We measured geocentric velocities of the target stars by cross-correlating target and template spectra
in the 2.29$~\micron$ to 2.31$~\micron$ region. Figure~\ref{fig-irs7full} shows a full spectrum
of IRS7 for observed orders, and
Figures~\ref{fig-bothirs7xcor} through \ref{fig-IRS20_gc1high} show
sample spectra of target stars
compared to the rest spectrum of Arcturus in the region covering the CO bandhead.\footnote{A complete
set of such figures for all stars in this paper can be found at: http://www.stsci.edu/~figer/papers.html.}
The spectra have been shifted along the
wavelength scale to maximize the cross-correlated power between the
target and template spectra.

We adjusted the geocentric velocities to the heliocentric frame by removing Earth motion,
+6.6~\kms~in June and $-$8.1~\kms~in July, where a positive value indicates motion toward the GC.
The error in the resultant velocity due to this adjustment is estimated to be $\pm$0.10~\kms.

A correction to the velocities was also made to account for the differences
in the spectral types of target stars versus that of Arcturus (K1.5III).
This correction was determined in a two-step process. First, we
cross-correlated template spectra of giants in the Wallace-Hinkle (WH) atlas \citep{wal96} with that of
Arcturus, and assigned a velocity correction as a function of temperature, as shown
in Figure~\ref{fig-vcorr_vs_temperature}a. We removed data for variable stars (open circles
in the plot) because their measured velocities can be influenced by gas motion in their 
atmospheres. 
Second, we determined velocity corrections for the target stars based on the depths of the CO feautures
in the spectra; the relation between the CO equivalent width, EW$_{\rm CO}$,
and temperature is shown in Figure~\ref{fig-vcorr_vs_temperature}b.
The velocity offsets and temperatures for the template stars
were taken from \citet{ram98}.
For the equivalent width measurements, we defined the continuum as the average flux from
2.288~\micron-2.293~\micron, and we isolated the region from
2.294~\micron-2.300~\micron\ for measuring EW$_{\rm CO}$.
Note that this process produces a velocity correction of nearly zero
for a K1.5III star, i.e.\ Arcturus, as expected.
The error in the velocity correction has contributions from the error in
the measurement of EW$_{\rm CO}$, the error in the relation between temperature
and EW$_{\rm CO}$, and the error in the relation between the
velocity correction and temperature. We assign an error of 2~$\AA$ to
the EW$_{\rm CO}$, or 166~K in temperature. The error in the temperature versus EW$_{\rm CO}$ relation is 81~K. Together,
these two errors contribute 0.8~\kms\ error in the velocity correction.
Summing this, in quadrature, with the error in the velocity correction versus temperature relation gives a total error
of 1.39~\kms. There is another potential source in error if the target stars
are long-period variables. In that case, the outer atmospheres exhibit variable
radial motions with respect to the systemic velocities of the stars. This might be
an important consideration for the latest types in the target sample. 

In summary, we estimate the following systematic errors in determining heliocentric stellar
velocities: Arcturus rest frame adjustment ($\pm$0.30~\kms), Earth motion adjustment ($\pm$0.10~\kms),
and spectral type correction ($\pm$1.39~\kms).
The total systematic error is then $\pm$1.43~\kms; note that the relative velocity errors are
smaller and can be determined from the data (see below).

Table~4 lists identifiers, relative positions, estimated spectral types, heliocentric velocities,
and internal velocity errors for each cool star, where
$V_{\star} = V_{\rm measured} + V_{\earth} - V_{\rm corr}$.

\section{Analysis and Results}
In this section, we present an anlysis of the statistics concerning
the velocity distribution and associated errors.

\subsection{Statistics and Errors}
Table~5 gives statistics of the measurements presented in Table~4. In particular,
we measured velocities for 78 stars in the first data set, and 29 stars in
the second data set. There are 22 stars in common in both data sets, and there
are 85 unique stars in the combined data set. Figure~\ref{fig-gchist} shows the histogram
of velocities from the combined data set, along with a similar plot for data
from \citet{gen00}. Figure~\ref{fig-gcbub} shows a ``bubble plot''
of stars and their velocities from the combined data set.
The unweighted average ($-$10.1$\pm$11.0~\kms) and
median ($-$7.5$\pm$11.0~\kms) velocities for the combined data set are consistent with each other and
consistent with the expected value resulting from the projection
of the Sun's peculiar velocity with respect to the Local Standard of
Rest (LSR), $-10\pm0.36$~\kms\ \citep{deh98};
this agreement suggests that there are no large systematic errors in
the absolute velocity calibration. Strictly speaking, however, it is
possible that the target population could exhibit bulk motion along the
line of sight with respect to the LSR \citep{mil96}.

The mean difference in velocities, in cases where
we measured two velocities for a single star, is $-$0.80~\kms, with a standard
deviation of 1.53~\kms. We assume that the individual errors are uncorrelated,
so that the 1$\sigma$ velocity error for cases with two velocity measurements, is 0.77~\kms,
and 1.08~\kms\ for cases with one velocity measurement.
Thus, the errors in Table~4 are small, and the velocities are consistent
with those in \citet{gen00}, as seen in Figures~\ref{fig-gchist} and \ref{fig-gcvsgen}. The standard deviation of velocities
for our combined data set is 100.9$\pm$7.7~\kms, where we take the error to be dominated
by statistics, i.e.\ error=$\sigma/(2N)^{0.5}$.

\subsection{Asymmetry}

We examined the data for asymmetric distributions in order to investigate the dynamical
state of stars in the sample. To first order, the velocity distribution
appears to be Gaussian (Figure~\ref{fig-gchist}). To inspect the velocity distribution further,
we examined the data for rotation about axes along galactic longitude and latitude.
Figure~\ref{fig-lv} shows the line-of-sight velocities ($v$) as a function
of the Galactic longitude and latitude (the Sun-GC distance is assumed
to be 8~{\it kpc}). It appears that there is no significant correlation
between velocity and projected distance from Sgr~A*, implying that the 85 sample stars
do not show any considerable figure rotation.
To explore this question in a more statistically robust way,
we examined the observed velocity distribution for rotation
about an arbitrary axis in the plane of the sky. To do this, we defined an asymmetry statistic,
\begin{equation} \label{equation-delta}
\Delta= \left[ n_{\rm red} - n_{\rm blue} \right]_{\rm \alpha_{\rm axis}
< PA < \alpha_{\rm axis}+180\arcdeg} - \left[ n_{\rm red} - n_{\rm blue} \right]_{\rm \alpha_{\rm axis}+180 < PA < \alpha_{\rm axis}+360\arcdeg},
\end{equation}
where ``red'' and ``blue'' refer to redshifted and blueshifted velocities,
and the terms in the brackets refer to the number of stars with position
angles between the angles given in the subscripts, where
$\alpha_{\rm axis}$ is the position of an assumed axis.
The position angles are measured
East of North, for an axis with the origin at the location of Sgr~A*.

The normalized asymmetry statistic is shown in Figure~\ref{fig-gcrot_ours}a for all possible assumed position angles
for the axis. The solid line has a maximum value at a position angle of 0$\arcdeg$,
a somewhat surprising, and perhaps coincidental, result. Otherwise, there
appears to be no significant asymmetry by this statistic. We performed Monte Carlo
simulations in order to determine the significance of the peak in $\Delta$.

Figure~\ref{fig-gcrot_ours}b shows the cumulative normalized probability of a model system having
a normalized asymmetry statistic, $\Delta$/N,
greater than the value on the x-axis. Model systems have 85 stars, each
with uniform random locations within a 1 by 1~{\it pc} box centered on
Sgr~A* and normal random velocities with $\sigma$=101~\kms. One thousand
systems were generated, and $\Delta$ was determined for position angles
between zero and 180$\arcdeg$ at 1.5$\arcdeg$ increments. The maximum
$\Delta$ was then stored for each system. The curve shows that approximately
20\% of systems will have $\Delta$ at least as great as 0.25, the value
we observe. This suggests that the observed system is consistent with
a symmetric distribution in the asymmetry statistic to within approximately 1$\sigma$.

\section{Discussion}

In this section, we discuss the results and their implications for determining
the dynamical properties of the late-type population in the GC.
We also describe their correspondence with the data from the 198 cool stars in \citet{gen00}, 
106 of which are located with the region covered by our observations, and 
49 of which are common with stars in our data set.
In addition, we discuss the possibility of using the data as
a first epoch of observations that will yield measurements of orbital accelerations
upon combining data from future observations.

Our results confirm that the sample population traces a distribution of
stars in isotropic orbits around a large dark mass.
While a few of the brightest stars in our sample are young, i.e.\ IRS7, they
evidently do not influence the statistics enough to imprint on them any velocity
asymmetry in the statistics we have examined.

\subsection{Comparison to Radial Velocities in \citet{gen00}}

There are 49 stars in common between our data set and that in \citet{gen00}.
For a detailed comparison between the two data sets, we have plotted the normalized asymmetry statistic and associated probability
function in Figures~\ref{fig-gcrot_genzel}a and \ref{fig-gcrot_genzel}b from data in \citet{gen00}. In general, the
asymmetry curves in Figures~\ref{fig-gcrot_genzel}a and \ref{fig-gcrot_ours}a are similar, showing maxima near
0$\arcdeg$, and two other broad humps at larger angles. Curiously, the
\citet{gen00} data show a conspicuous number of stars with positive
velocities on either side of a rotation axis, regardless of position angle,
as seen in the dotted and dashed lines. This is consistent with the
fact that the median and mean velocity in their data set are +12.0~\kms\ and +2.2~\kms, with an error of 7.3~\kms, as opposed
to $-$10.1~\kms\ and $-$7.5~\kms, with an error of 11.0~\kms, in ours; both data sets are consistent with
a Gaussian velocity distribution having similar standard deviations, 100.9$\pm$7.7~\kms\ in our data
and 103$\pm$5.2~\kms\ in theirs.

\subsection{Enclosed Mass Estimates}

Both data sets also produce similar results when used to infer an
enclosed mass, assuming a relationship
between the velocity dispersion and the local gravitational potential, i.e.\
the Bahcall-Tremaine \citep{bah81} estimator:
\begin{equation} \label{equation-BT}
M_{BT}={{16} \over {\pi G N}} \sum_{R<R_0}^{} v^2 R w,
\end{equation}
where, G is the gravitational constant, N is the number of stars,
v is the line-of-sight velocity, R is the projected radius, w is the normalized statistical
weight, and the sum is over all test particles within the enclosed projected radius, R$_{\rm 0}$.
The individual weights are assigned to be one over the square of
the errors in Table~4, i.e.\ the inverse variances, and the mean velocity of
the sample is subtracted before applying Equation~\ref{equation-BT}.


Figures~\ref{fig-vdisp}a, \ref{fig-vdisp}b, \ref{fig-mass}a, and \ref{fig-mass}b show the velocity dispersion, $\sigma$, and
enclosed mass, {\it M}$_{\rm BT}$, as calculated using data in Table~4 ({\it a}) and from \citet{gen00} ({\it b});
we shifted velocities in both data sets to adjust the mean of the sample
to zero. The error bars are the 1$\sigma$ values calculated from the data.
The \citet{gen00} data extend to smaller radii, but both data sets show the
trend of a few million solar masses of enclosed material at the
smallest radius, and a steady increase of enclosed mass for larger radii.
Within errors, the data sets agree in their prediction of enclosed mass.
Because it is tempting to over-interpret
these figures, we limit our conclusions to be that the enclosed mass within a projected
radius of 0.1~{\it pc} is $\sim$2$\pm$1(10$^6$)~\Msun\ according to the BT statistic.
This estimate compares favorably with others \citep{lac82,ser85,mcg89,sel90,hal96a,gen97,ghe98,gen00,ghe00,eck02},
although they tend to be low with respect to the ``true'' enclosed mass, 3.45$\pm$0.5(10$^6$)~\Msun,
inferred by assuming a Keplerian orbit for one of the closest stars to
the dark mass \citep{sch02,ghe03}.

One notable weakness of the BT estimator in this application is the fact
that the distribution of observed particles may deviate from that assumed
in the derivation of the estimator. In particular, our
coverage represents a pencil-beam that slices through the GC,
thereby including test points significantly farther from the
Center than the projected extent of our sample in the plane of the sky; this
will tend to add objects with lower line-of-sight velocities than would be
the case if we were completely sampling all projeced radii in the system. In
addition, our sample includes stars on plunging orbits, i.e.\ with
high line-of-sight velocities, without containing the corresponding group
of stars from the isotropic distribution at projected radii beyond what
was sampled. This bias will tend to increase the estimated projected mass.
\citet{hal96b} highlight these biases and conclude that one could
not likely estimate the enclosed mass in the central parsec to better than
a factor of two using the BT estimator. An additional bias can be introduced
if particles are missing from the distribution, i.e.\ in the case of an evacuated
cavity, as we shall see later.

\subsection{Models}

We attempted to fit the radial profiles and
velocity dispersions assuming a relatively flat distribution, within some radius, r$_0$,
that is described by a power law of index n, according to equation~\ref{equation-density}, and
similar to the form used by \citet{sel90}.

\begin{equation} \label{equation-density}
\rho  \propto
\left\{
\begin{array}{l}
r^n, \mbox{for r$<$r$_{0}$};\\
r^{-2}, \mbox{for r$>$r$_{0}$}.
\end{array}
\right.
\end{equation}

We find good agreement between the models and observations for r$_0\sim$0.4~\pc, and n$\sim-$1/4,
using the following procedure.
We created a set of model test points in a Monte Carlo simulation assuming equation~\ref{equation-density} from 0 to 10~\pc,
and a total enclosed mass at the outer radius of 3(10$^7$)~\Msun\ \citep{mcg89}.
Each particle is governed by a gravitational potential set by the
the enclosed mass within the radius of the particle, including the masses of particles with orbits
at smaller radii and a central point mass to represent the black hole, having
M$_{\rm BH}$=3.45(10$^6$)~\Msun, the mean of the values in \citet{sch03} and \citet{ghe03}. Each particle
is on a circular orbit, with uniform random phase and inclination with respect
to the line of sight. The total sample size $\sim$10$^6$ particles. We
used the simulation to find the most probable r$_0$ of $\sim$0.4~\pc.
We then determined the best-fit index by minimizing
the $\chi^2$ of the difference between the model and observed
velocity dispersion (Figures~\ref{fig-vdisp_model_n=-0.25_ours_r0=0.4}a and \ref{fig-vdisp_model_n=-0.25_ours_r0=0.4}b).
The results are shown in Figures~\ref{fig-chisq_ours_r0=0.4}a and \ref{fig-chisq_ours_r0=0.4}b, where we see that the best
fit is for a power law index $\sim-$1/4 for r$<$r$_0$.

Figures~\ref{fig-model_0.4circ.xyproj}a and \ref{fig-model_0.4circ.xyproj}b show plots of locations for
10,000 randomly selected points in a simulation with n=$-2$ ({\it a}) and one with
n=$-$1/4 ({\it b}).
Figures~\ref{fig-model_0.0circ}a and \ref{fig-model_0.0circ}b
show the projected densities and velocity dispersions for the modelled particles, for circular
and isotropic orbits.
The surface density increases as 1/r outside or r$_0$, as expected, and rolls
over to a roughly constant value within this radius. The resultant velocity
dispersion is relatively flat within r$_0$, consistent with the
observed behavior.

\citet{sel90} and \citet{hal96a} describe the absence of the CO bandhead feature in spectra
of integrated light within 0.6~\pc\ of the GC. They consider the hypothesis
that CO in the atmospheres of late-type stars is destroyed by photo-dissociation or stellar collisions
in the region. Others
have also suggested yet other explanations for the CO hole, i.e.\
stellar mergers/collisions \citep{lac82,lee96,ale99,bai99},
atmospheric stripping \citep{ale01}, and atmospheric heating \citep{ale03}.
\citet{gen96} provide strong evidence for a hole of late-type stars in the form
of a steep decrease in the projected number density of such stars (their Figure~6) and
a decrease in the strength of the CO bandhead feature in integrated light within 5\arcsec\
of the GC. They also invoke these features as evidence for the destruction of the largest
red giants in interactions with main sequence stars.
Our findings are consistent with previously obtained evidence for a flattening of
the density distribution; however,
we consider that yet another physical process may explain the observed phenomenon.
\citet{gou00} argue that a cluster of $\sim$24,000 stellar mass black
holes should have migrated toward the central parsec through dynamical
friction on a timescale of $\sim$10~\Gyr\ \citep{mor93}. They find
through simulations that this cluster will eject stars on a time scale of $\sim$1~\Gyr.
We propose that such a cluster of black holes could provide the scattering
sources for ejecting relatively low mass, and old, stars from the region.

We favor this hypothesis versus the alternatives for the following reasons.
First, the requisite star formation needed to produce the stellar black holes is supported
by the presence of young and intermediate-age stars in the GC, i.e.\ \citet{leb82}, \citet{rie87},
\citet{leb87}, \citet{hal89}, \citet{hal92}, \citet{rie93}, \citet{blu95},
\citet{kra95}, \citet{gen96}, \citet{nar96}, \citet{sjo99}, and \citet{fig03}.
Second, \citet{mor93} and \citet{gou00}
predict that dynamical friction should bring black hole remnants from massive
stars into the central parsec in substantially less than 10~\Gyr.
Alternate hypotheses require interaction (mergers or collisions) rates that may be
in excess of what is predicted by the observed stellar density distribution and/or
stellar evolution scenarios that are poorly understood, i.e.\ the spectral energy
distribution emitted by a post-merger star; however, see \citet{gen03} for
an alternate interpretation.

Finally, we explore the difference in enclosed mass estimates from the
proper motion studies, M$\sim$2.5$-$3.2(10$^6$)~\Msun\ \citep{gen96,ghe98}, and those
from more recent estimates derived from single orbits, M=3.45$\pm$1.5(10$^6$)~\Msun\ \citep{sch03,ghe03}.
The former are inferred using a mass estimator, such as the BT estimator.
We attribute the difference as due to differences in the observed distribution
versus that assumed in the derivation of the BT estimator. To take an example,
consider Figures~\ref{fig-model_0.4circ.mass}a and \ref{fig-model_0.4circ.mass}b, in which we plot the true enclosed
mass and that estimated by the BT estimator, versus radius, for a system with no hole ({\it a}),
and with a hole having r$_0$=0.4~\pc\ ({\it b}). In this case, we are
using circular orbits, so the normalization factor in equation~\ref{equation-BT}
is 32/3, instead of 16. The plot shows that the BT estimator overestimates the enclosed
mass at large radii, as expected from an incompletely populated system, and underestimates
the enclosed mass at radii less than the hole radius. Note that the true mass and that
inferred using the BT estimator converge for radii above r$_0$ up to $\sim$1~\pc.
This effect is also seen in the observations (Figure~\ref{fig-mass}) where the enclosed
mass estimates converge to $\sim$3.8(10$^6$)~\Msun\ for a projected radius above 0.4~\pc, in
good agreement with the black hole mass as inferred from single orbits.

We have used the models in this section to explore the plausibility that a
hole in the distribution of old stars can reproduce the observed velocity dispersion and
projected number distributions. While the models are consistent with the presence
of such a hole, we look forward to more robust modeling that can be tested
with statistical methods to determine the most likely volume density of
old stars in the GC. If a hole is required, then it will be interesting
to determine the physical mechanism that produces it.
In particular, it will be interesting to determine whether the Chandra sources in the central
parsec are consistent with the presence of a cluster of black holes \citep{mun02}.

\subsection{Accelerations}
We expect to be able to measure accelerations by comparing velocities in this paper
to those collected in the future. The magnitudes of these accelerations depend on the
projected radius and orbital inclination, as shown in Figure~\ref{fig-accel}, where
the potential is defined by a point mass having M=3.45(10$^6$)~\Msun.
For an ``average'' case,
we expect to measure an acceleration that exceeds the errors in 5 years from the
first epoch of observations in this paper. Note that the error in the estimated acceleration
only depends on the statistical errors of the measurements, and not the systematic
errors, i.e.\ the peculiar solar velocity with respect to the local standard of rest.

\section{Conclusions}
We have established a first epoch of radial velocity data for a large (85) number
of cool stars in the central parsec of the Galaxy. The data set have much higher
precision, by at least a factor of 30, than those previously obtained for similar
samples. The data agree with previous results, and we conclude that the late-type
population has Gaussian velocities in the line-of-sight component and that the
enclosed mass at a projected radius of 0.2~{\it pc} is consistent with a value
$\sim$3.45(10$^6$)~\Msun.
We have identified dynamical evidence for a flattened distribution of late-type stars within the
central parsec, with an edge at $\sim$0.4~\pc.
The finding adds evidence of such a flattening, in addition to that
from previous studies finding an absence of the CO bandhead feature in the
integrated light of late-type stars in the region. While the integrated light measurements
might suffer from dilution by the light of young stars, the dynamical evidence
does not suffer from such a bias.

\acknowledgements
We acknowledge useful conversations with Mike Fall. We also acknowledge
the work of: Maryanne Angliongto, Oddvar
Bendiksen, George Brims, Leah Buchholz, John Canfield, Kim Chin, Jonah
Hare, Fred Lacayanga, Samuel B. Larson, Tim Liu, Nick Magnone, Gunnar
Skulason, Michael Spencer, Jason Weiss and Woon Wong. In addition, we
thank the CARA instrument specialist Thomas A. Bida, and all the CARA staff
involved in the commissioning and integration of NIRSPEC.

\clearpage
\begin{deluxetable}{rcrrrcr}
\small
\tablewidth{0pt}
\tablecaption{Log of Observations}
\tablehead{
\colhead{Name\tablenotemark{a}} &
\colhead{Resolution\tablenotemark{b}} &
\colhead{Filter\tablenotemark{c}} &
\colhead{Integ.} &
\colhead{Frames} &
\colhead{Slit Size} &
\colhead{Date} }
\startdata
GC1 & 14,000 & NIRSPEC-7 & 60 s. & 40 & 0$\farcs72\times 24\arcsec$ & 4 June 1999 \\
GC2 & 23,300 & NIRSPEC-6/7 & 150 s. & 67 & 0\farcs$43\times 24$\arcsec & 4 July 1999 \\
\enddata
\tablenotetext{a}{``GC1'' refers to the first slit scan while ``GC2'' refers
to the second slit scan.}
\tablenotetext{b}{The resolution is $\lambda$/$\Delta\lambda_{\rm FWHM}$, where
$\Delta\lambda_{\rm FWHM}$ is the half-power line width of unresolved arc lamp lines.
The slit width was 5 pixels in GC1 and 3 pixels in GC2.}
\tablenotetext{c}{NIRSPEC-7 has half-power points of 1.85~\micron\ and
2.62~\micron. NIRSPEC-6 has half-power points of 1.56~\micron\ and
2.30~\micron. Half-power points are from \citet{fig99c}.
Because the orders are longer than the width of the
detector, the spectra are non-contiguous in wavelength.}
\tablecomments{All images were obtained with the slit positioned approximately north-south.
The multiple correlated read mode \citep{fow90} with 16 reads at the beginning and
end of each integration was used for both data sets.}
\end{deluxetable}

\clearpage
\begin{deluxetable}{crccrcc}
\small
\tablewidth{0pt}
\tablecaption{Wavelength Coverage of Echelle Orders ($\micron$)}
\tablehead{
\colhead{Echelle} &
\colhead{} &
\colhead{GC1\tablenotemark{a}} &
\colhead{GC1\tablenotemark{a}} &
\colhead{} &
\colhead{GC2} &
\colhead{GC2} \\
\colhead{Order} &
\colhead{} &
\colhead{NIRSPEC-7} &
\colhead{NIRSPEC-7} &
\colhead{} &
\colhead{NIRSPEC-6/7} &
\colhead{NIRSPEC-6/7} \\
\colhead{} &
\colhead{} &
\colhead{min} &
\colhead{max} &
\colhead{} &
\colhead{min} &
\colhead{max} }
\startdata
33  & &  2.281 & 2.315  & &   2.290 & 2.324 \\
34  & &  2.214 & 2.248  & &   2.223 & 2.256 \\
35  & &  2.152 & 2.184  & &   2.160 & 2.192 \\
36  & &  2.092 & 2.124  & &   2.101 & 2.131 \\
37  & &  2.036 & 2.067  & &   2.044 & 2.074 \\
38  & &  1.983 & 2.013  & &   1.991 & 2.020 \\
\enddata
\tablenotetext{a}{``GC1'' refers to the first slit scan while ``GC2'' refers
to the second slit scan.}
\end{deluxetable}

\clearpage
\begin{deluxetable}{rrrrr}
\small
\tablewidth{0pt}
\tablecaption{Velocity Correction for Spectral Type}
\tablehead{
\colhead{WH atlas\tablenotemark{a}} &
\colhead{V$_{\rm off}$\tablenotemark{b}} &
\colhead{Type} &
\colhead{T$_{\rm eff}$\tablenotemark{c}} &
\colhead{V$_{\rm corr}$\tablenotemark{d}} \\
\colhead{Star} &
\colhead{\kms} &
\colhead{} &
\colhead{K} &
\colhead{\kms}
}
\startdata
Kap Gem & 1.86 & G8III & 4866 & 2.31 \\
Xi Her & 3.93 & G8.5III & 4788 & 1.97 \\
Iot Cep & 1.60 & K0-III & 4565 & 1.01 \\
11 Cep & 0.46 & K0.5III & 4495 & 0.71 \\
Alp Boo & $-$0.64 & K1.5IIIp & 4360 & 0.13 \\
Kap Oph & $-$0.46 & K2III & 4294 & $-$0.15 \\
39 Cyg & $-$2.00 & K2.5III & 4232 & $-$0.42 \\
31 Lyn & $-$1.89 & K4.5III & 4000 & $-$1.42 \\
Gam Dra & $-$2.25 & K5III & 3946 & $-$1.65 \\
Alp Tau & $-$1.06 & K5+III & 3946 & $-$1.65 \\
Mu UMa & $-$4.79 & M0III & 3845 & $-$2.09 \\
Gam Sge & $-$1.17 & M0-III & 3845 & $-$2.09 \\
Nu Vir & $-$1.94 & M1III & 3752 & $-$2.49 \\
75 Cyg & $-$1.65 & M1IIIab & 3752 & $-$2.49 \\
 HR8989 & $-$1.61 & M2III & 3666 & $-$2.86 \\
Chi Peg & $-$2.42 & M2+III & 3666 & $-$2.86 \\
 HR8621 & $-$10.69 & M4+III & 3517 & \nodata \\
R Lyr & $-$4.34 & M5III & 3454 & \nodata \\
 HR8530 & $-$14.91 & M6III & 3399 & \nodata \\
EU Del & $-$4.14 & M6III & 3399 & \nodata \\
SW Vir & $-$4.85 & M7III: & 3351 & \nodata \\
BK Vir & $-$10.58 & M7-III: & 3351 & \nodata \\
RX Boo & $-$4.47 & M7.5-8III & 3320 & \nodata \\
\enddata
\tablenotetext{a}{Giants from the Wallace-Hinkle atlas \citep{wal96}.}
\tablenotetext{b}{Measured velocity offset with respect to Arcturus.}
\tablenotetext{c}{Temperatures from \citet{ram98}.}
\tablenotetext{d}{Fit velocity offset with respect to Arcturus.}
\end{deluxetable}

\clearpage
\begin{deluxetable}{clrrrcrrrrrr}
\tabletypesize{\tiny}
\tablewidth{470pt}
\tablecaption{Velocity Data}
\tablehead{
\colhead{Identifier} &
\colhead{Designation\tablenotemark{a}} &
\colhead{$\Delta$RA\tablenotemark{b}} &
\colhead{$\Delta$DEC\tablenotemark{b}} &
\colhead{r\tablenotemark{b}} &
\colhead{Sp.\tablenotemark{c}} &
\colhead{V$_1$\tablenotemark{d}} &
\colhead{V$_2$\tablenotemark{e}} &
\colhead{V$_{\rm AVE}$\tablenotemark{f}} &
\colhead{$\sigma$\tablenotemark{g}} &
\colhead{V$_{\rm Genzel}$\tablenotemark{h}} &
\colhead{$\sigma_{\rm Genzel}$\tablenotemark{h}} \\
\colhead{} &
\colhead{} &
\colhead{$\arcsec$} &
\colhead{$\arcsec$} &
\colhead{$\arcsec$} &
\colhead{Type} &
\colhead{\kms} &
\colhead{\kms} &
\colhead{\kms} &
\colhead{\kms} &
\colhead{\kms} &
\colhead{\kms} \\
\colhead{(1)} &
\colhead{(2)} &
\colhead{(3)} &
\colhead{(4)} &
\colhead{(5)} &
\colhead{(6)} &
\colhead{(7)} &
\colhead{(8)} &
\colhead{(9)} &
\colhead{(10)} &
\colhead{(11)} &
\colhead{(12)}
}
\startdata
259 & \nodata & $-$0.68 & $-$1.72 & 1.85 & K4III & 95.7 & 97.4 & 96.5 &   0.8 & \nodata & \nodata \\
313 & IRS29S & $-$1.79 & 0.82 & 1.97 & K3III & $-$160.3 & \nodata & $-$160.3 &   1.1 & $-$93 &   20 \\
327 & \nodata & $-$1.58 & 1.24 & 2.00 & G9III & $-$156.0 & \nodata & $-$156.0 &   1.1 & \nodata & \nodata \\
360 & G288? & $-$1.28 & 2.61 & 2.91 & G9III & \nodata & 133.7 & 133.7 &   1.1 & 17 &   30 \\
365 & \nodata & 1.23 & 2.79 & 3.05 & K2III & 1.9 & \nodata & 1.9 &   1.1 & \nodata & \nodata \\
221 & \nodata & 0.99 & $-$3.36 & 3.50 & M6III & $-$16.0 & \nodata & $-$16.0 &   1.1 & \nodata & \nodata \\
222 & IRS33W & $-$0.03 & $-$3.51 & 3.51 & K4III & 72.0 & 75.0 & 73.5 &   0.8 & 82 &   25 \\
242 & IRS13W & $-$3.92 & $-$2.51 & 4.66 & K3III & $-$40.0 & \nodata & $-$40.0 &   1.1 & $-$74 &   30 \\
244 & \nodata & 4.09 & $-$2.35 & 4.72 & M0III & 40.9 & 39.7 & 40.3 &   0.8 & \nodata & \nodata \\
398 & \nodata & 1.41 & 4.76 & 4.97 & M0III & $-$262.8 & \nodata & $-$262.8 &   1.1 & \nodata & \nodata \\
416 & IRS7 & $-$0.26 & 5.65 & 5.65 & $>$M7III\tablenotemark{i} & $-$108.6 & $-$109.4 & $-$109.0 &   0.8 & $-$103 &   15 \\
170 & IRS20 & $-$0.40 & $-$5.68 & 5.69 & M2III & $-$90.7 & $-$91.1 & $-$90.9 &   0.8 & 17 &   25 \\
153 & G577A? & 0.19 & $-$6.45 & 6.46 & M1III & 27.2 & 29.8 & 28.5 &   0.8 & 32 &   25 \\
418 & \nodata & $-$2.73 & 5.92 & 6.52 & K3III & $-$115.5 & \nodata & $-$115.5 &   1.1 & \nodata & \nodata \\
189 & \nodata & 4.72 & $-$4.81 & 6.74 & K2III & \nodata & 26.9 & 26.9 &   1.1 & \nodata & \nodata \\
308 & \nodata & 6.84 & 0.46 & 6.85 & K0III & 63.1 & \nodata & 63.1 &   1.1 & \nodata & \nodata \\
213 & \nodata & $-$5.74 & $-$3.78 & 6.87 & K1III & 41.1 & \nodata & 41.1 &   1.1 & \nodata & \nodata \\
144 & \nodata & $-$1.20 & $-$6.96 & 7.06 & M0III & $-$23.4 & $-$21.2 & $-$22.3 &   0.8 & \nodata & \nodata \\
415 & \nodata & 4.45 & 5.63 & 7.17 & K1III & 68.7 & 67.7 & 68.2 &   0.8 & \nodata & \nodata \\
430 & \nodata & 2.99 & 6.57 & 7.22 & G8III & \nodata & 16.0 & 16.0 &   1.1 & \nodata & \nodata \\
325 & IRS1NE(3) & 7.34 & 1.23 & 7.44 & M1III & \nodata & 72.7 & 72.7 &   1.1 & 29 &   25 \\
297 & \nodata & 7.48 & 0.06 & 7.48 & G9III & \nodata & 71.1 & 71.1 &   1.1 & \nodata & \nodata \\
346 & IRS1NE(2) & 7.42 & 2.20 & 7.74 & M5III & \nodata & $-$68.8 & $-$68.8 &   1.1 & $-$8 &   25 \\
126 & \nodata & 0.07 & $-$7.81 & 7.81 & M0III & $-$145.3 & $-$147.0 & $-$146.2 &   0.8 & \nodata & \nodata \\
137 & IRS12N & $-$2.81 & $-$7.50 & 8.01 & M0III & $-$60.4 & $-$60.9 & $-$60.7 &   0.8 & $-$96 &   20 \\
407 & BHA4E & $-$6.08 & 5.36 & 8.11 & M4III & $-$28.2 & \nodata & $-$28.2 &   1.1 & $-$77 &   30 \\
459 & G734? & $-$3.16 & 7.72 & 8.34 & K3III & $-$28.3 & \nodata & $-$28.3 &   1.1 & 104 &   40 \\
342 & IRS1NE(1) & 8.12 & 1.99 & 8.36 & M3III & $-$67.5 & \nodata & $-$67.5 &   1.1 & 112 &   25 \\
124 & G849? & 3.30 & $-$8.02 & 8.67 & K2III & $-$107.3 & \nodata & $-$107.3 &   1.1 & 88 &   25 \\
115 & IRS14N & 1.58 & $-$8.56 & 8.70 & M4III & $-$13.9 & $-$13.3 & $-$13.6 &   0.8 & 19 &   20 \\
417 & BHA4W & $-$6.89 & 5.65 & 8.91 & K2III & 88.4 & \nodata & 88.4 &   1.1 & 100 &   30 \\
467 & \nodata & $-$2.62 & 8.56 & 8.95 & M4III & $-$15.5 & $-$12.1 & $-$13.8 &   0.8 & \nodata & \nodata \\
151 & IRS9 & 6.47 & $-$6.42 & 9.11 & M3III & $-$341.4 & $-$341.3 & $-$341.3 &   0.8 & $-$300 &   25 \\
96 & IRS14SW & 0.41 & $-$9.47 & 9.48 & M5III & 24.5 & 24.6 & 24.5 &   0.8 & 29 &   20 \\
493 & G904 & 0.27 & 9.51 & 9.51 & M2III & 43.4 & 44.2 & 43.8 &   0.8 & 23 &   30 \\
101 & IRS12S & $-$2.87 & $-$9.26 & 9.70 & M1III & 48.9 & 48.7 & 48.8 &   0.8 & 51 &   25 \\
271 & G986 & $-$9.63 & $-$1.25 & 9.71 & M0III & 62.7 & \nodata & 62.7 &   1.1 & $-$50 &   30 \\
87 & \nodata & $-$1.23 & $-$9.76 & 9.84 & M1III & $-$46.9 & $-$45.1 & $-$46.0 &   0.8 & \nodata & \nodata \\
498 & \nodata & $-$0.95 & 10.01 & 10.06 & K5III & $-$5.2 & $-$1.9 & $-$3.5 &   0.8 & \nodata & \nodata \\
393 & IRS10EE & 8.94 & 4.65 & 10.08 & M0III & $-$113.1 & \nodata & $-$113.1 &   1.1 & $-$55 &   30 \\
290 & \nodata & $-$10.19 & $-$0.28 & 10.20 & M2III & $-$25.9 & \nodata & $-$25.9 &   1.1 & \nodata & \nodata \\
94 & \nodata & $-$3.80 & $-$9.60 & 10.32 & K4III & $-$7.5 & \nodata & $-$7.5 &   1.1 & \nodata & \nodata \\
174 & G1060? & 8.89 & $-$5.44 & 10.42 & K1III & 25.6 & \nodata & 25.6 &   1.1 & $-$103 &   35 \\
495 & G1044 & $-$3.89 & 9.75 & 10.50 & M2III & 199.5 & 200.3 & 199.9 &   0.8 & 134 &   25 \\
460 & \nodata & $-$8.07 & 7.76 & 11.20 & M0III & $-$100.6 & \nodata & $-$100.6 &   1.1 & \nodata & \nodata \\
490 & G1130 & $-$6.25 & 9.35 & 11.24 & M3III & $-$20.6 & \nodata & $-$20.6 &   1.1 & 7 &   30 \\
338 & G1061? & 11.14 & 1.77 & 11.28 & M1III & 184.8 & \nodata & 184.8 &   1.1 & 227 &   25 \\
410 & \nodata & 9.92 & 5.48 & 11.33 & K2III & $-$107.1 & \nodata & $-$107.1 &   1.1 & \nodata & \nodata \\
171 & \nodata & $-$9.90 & $-$5.56 & 11.35 & K3III & 24.5 & \nodata & 24.5 &   1.1 & \nodata & \nodata \\
134 & \nodata & 8.57 & $-$7.51 & 11.39 & G9III & $-$9.3 & \nodata & $-$9.3 &   1.1 & \nodata & \nodata \\
437 & G1150 & $-$9.10 & 6.87 & 11.40 & M4III & 15.6 & \nodata & 15.6 &   1.1 & 59 &   30 \\
366 & \nodata & $-$11.03 & 3.02 & 11.43 & K4III & $-$32.9 & \nodata & $-$32.9 &   1.1 & \nodata & \nodata \\
534 & \nodata & 2.02 & 11.63 & 11.81 & K3III & $-$240.6 & \nodata & $-$240.6 &   1.1 & \nodata & \nodata \\
538 & IRS15N & 0.83 & 11.78 & 11.81 & M2III & $-$12.3 & $-$10.1 & $-$11.2 &   0.8 & 17 &   25 \\
527 & G1152 & $-$3.42 & 11.47 & 11.97 & M3III & $-$43.5 & \nodata & $-$43.5 &   1.1 & $-$68 &   30 \\
68 & \nodata & $-$5.35 & $-$10.71 & 11.97 & M0III & $-$64.8 & \nodata & $-$64.8 &   1.1 & \nodata & \nodata \\
441 & G1174? & 9.70 & 7.15 & 12.05 & K1III & $-$0.1 & \nodata & $-$0.1 &   1.1 & 12 &   50 \\
295 & G1124? & 12.06 & 0.00 & 12.06 & $<$G8III & 224.8 & \nodata & 224.8 &   1.1 & 177 &   30 \\
71 & G1248 & 6.45 & $-$10.23 & 12.09 & $<$G8III & \nodata & 80.6 & 80.6 &   1.1 & 171 &   35 \\
474 & G1118? & 8.26 & 8.88 & 12.13 & K1III & 47.9 & \nodata & 47.9 &   1.1 & 105 &   35 \\
541 & G1183 & $-$1.10 & 12.09 & 12.14 & M2III & $-$4.6 & \nodata & $-$4.6 &   1.1 & 45 &   20 \\
404 & \nodata & $-$11.16 & 5.10 & 12.28 & M0III & $-$77.1 & \nodata & $-$77.1 &   1.1 & \nodata & \nodata \\
464 & G1266 & $-$9.30 & 8.10 & 12.33 & K4III & $-$113.7 & \nodata & $-$113.7 &   1.1 & 42 &   50 \\
524 & G1198 & $-$4.98 & 11.31 & 12.36 & M2III & $-$133.1 & \nodata & $-$133.1 &   1.1 & $-$88 &   50 \\
491 & \nodata & $-$8.19 & 9.38 & 12.45 & K3III & 38.2 & \nodata & 38.2 &   1.1 & \nodata & \nodata \\
261 & \nodata & 12.57 & $-$1.57 & 12.67 & G9III & 158.8 & \nodata & 158.8 &   1.1 & \nodata & \nodata \\
173 & IRS28 & 11.48 & $-$5.54 & 12.74 & M1III & $-$47.6 & \nodata & $-$47.6 &   1.1 & $-$93 &   25 \\
392 & G1202? & 12.05 & 4.51 & 12.86 & M3III & 79.9 & \nodata & 79.9 &   1.1 & 137 &   25 \\
516 & G1187? & 7.05 & 10.81 & 12.90 & M0III & 70.4 & 71.6 & 71.0 &   0.8 & 37 &   30 \\
528 & G1306 & $-$6.78 & 11.41 & 13.27 & M0III & 11.2 & \nodata & 11.2 &   1.1 & $-$23 &   50 \\
547 & G1311 & 5.59 & 12.33 & 13.54 & K4III & 84.7 & 84.4 & 84.5 &   0.8 & 111 &   30 \\
158 & G1390? & 12.18 & $-$6.32 & 13.72 & K4III & $-$218.6 & \nodata & $-$218.6 &   1.1 & $-$163 &   30 \\
102 & G1438? & $-$10.41 & $-$9.10 & 13.83 & M4III & 53.9 & \nodata & 53.9 &   1.1 & 72 &   25 \\
431 & G1271? & 12.28 & 6.77 & 14.03 & K1III & $-$53.3 & \nodata & $-$53.3 &   1.1 & 114 &   25 \\
506 & \nodata & $-$9.84 & 10.18 & 14.16 & K2III & $-$27.7 & \nodata & $-$27.7 &   1.1 & \nodata & \nodata \\
482 & G1458 & $-$10.98 & 9.12 & 14.27 & M4III & 36.8 & \nodata & 36.8 &   1.1 & 77 &   45 \\
462 & G1334? & 12.26 & 7.96 & 14.62 & K2III & $-$144.4 & \nodata & $-$144.4 &   1.1 & 67 &   40 \\
109 & \nodata & 12.56 & $-$8.67 & 15.26 & M0III & $-$17.9 & \nodata & $-$17.9 &   1.1 & \nodata & \nodata \\
570 & \nodata & 7.24 & 13.60 & 15.40 & K2III & $-$38.3 & \nodata & $-$38.3 &   1.1 & \nodata & \nodata \\
539 & IRS11SW & $-$10.39 & 11.80 & 15.72 & M6III & $-$91.6 & \nodata & $-$91.6 &   1.1 & $-$24 &   30 \\
81 & G1620? & 12.54 & $-$9.86 & 15.95 & G8III & 78.5 & \nodata & 78.5 &   1.1 & 81 &   35 \\
575 & \nodata & 8.16 & 13.75 & 15.99 & K3III & 143.3 & \nodata & 143.3 &   1.1 & \nodata & \nodata \\
573 & IRS11NE & $-$9.31 & 13.56 & 16.45 & M6III & $-$2.7 & \nodata & $-$2.7 &   1.1 & 14 &   30 \\
566 & \nodata & 10.87 & 13.26 & 17.15 & M0III & 44.1 & \nodata & 44.1 &   1.1 & \nodata & \nodata \\
580 & \nodata & 11.57 & 13.86 & 18.05 & G9III & 182.8 & \nodata & 182.8 &   1.1 & \nodata & \nodata \\
\enddata
\tablenotetext{a
}{Designations are taken from \citet{gen00}. The GNNNN designations are formed by converting the radius in \citet{gen00} into a string without the decimal point, i.e. G288 is the star with r=2\farcs88, and G213B is the second star with r=2\farcs13.}
\tablenotetext{b
}{Positions are with respect to Sgr~A* and have an error of $\pm$0.1\arcsec. A negative offset indicates that an object is to the south or the west of Sgr~A*.}
\tablenotetext{c
}{Late-type spectral types were determined by comparing spectra to those of stars in the WH atlas \citep{wal96}.}
\tablenotetext{d
}{Heliocentric velocities measured from data in first slit-scan.}
\tablenotetext{e
}{Heliocentric velocities measured from data in second slit-scan.}
\tablenotetext{f}{(V$_1$+V$_2$)/2.}
\tablenotetext{g}{Estimated error in average velocity.}
\tablenotetext{h}{Heliocentric velocities from \citet{gen00}.}
\tablenotetext{i}{IRS7 has a spectral type of M2I, according to \citet{sel87}, based on the
depth of the CO absorption features, the relative lack of water absorption, and the
estimated luminosity. Our code only considers CO absorption depth and 
assigns a very late red giant spectral type.}
\tablecomments{Negative velocities indicate a blue-shifted spectrum.}
\end{deluxetable}

\clearpage
\begin{deluxetable}{lrrrr}
\footnotesize
\tablewidth{430pt}
\tablecaption{Velocity Statistics}
\tablehead{
\colhead{Statistic} &
\colhead{GC1} &
\colhead{GC2} &
\colhead{GC1 or GC2} &
\colhead{GC1 and GC2} \\
\colhead{(1)} &
\colhead{(2)} &
\colhead{(3)} &
\colhead{(4)} &
\colhead{(5)}
}
\startdata
number  & 153 & 68 &\nodata & \nodata \\
number of cool stars  & 78 &29 & 85 &22 \\
average velocity  & $-$15.37 &9.04 &$-$10.09 &$-$3.58 \\
median velocity  & $-$12.32 &26.86 &$-$7.48 &24.52 \\
standard deviation  & 102.18 & 100.44 & 100.88 &107.65 \\
average V$_{\rm GC1}$ $-$ V$_{\rm GC2}$  & \nodata & \nodata & \nodata &$-$0.80 \\
$\sigma_{V_{\rm GC1} - V_{\rm GC2}}$ & \nodata & \nodata & \nodata & 1.53 \\
$\sigma_{\rm V}$  & 1.08 &1.08 & \nodata &0.77 \\
\enddata
\tablecomments{Negative velocities indicate a blue-shifted spectrum. $V_{\rm GC1} - V_{\rm GC2}$ is the velocity difference for stars that are common in both data sets. The sample of cool stars only includes those for which high quality spectra were obtained.}
\end{deluxetable}

\clearpage

\begin{figure}
\epsscale{1.8}
\hspace{1.75in}
\hspace*{1.5in}
\vskip .0in
\caption{{\it (a)} An image taken with the slit-viewing camera, with synthetic slit apertures overlayed to correspond to slit positions used to obtain the first night of data. The slit apertures
cover 24\arcsec\ in length and 0\farcs72 in width. {\it (b)} The same for the second night of data, except that
the slit width is 0\farcs43. 
This figure can be found in the full version of the paper at http://www-int.stsci.edu/$\sim$figer/papers/nirspec/vel/ms.ps. \label{fig-gc1scam_slits}}
\end{figure}
\clearpage

\begin{figure}\epsscale{1.3}
\hspace{3.75in}
\hspace*{4.5in}
\vskip .2in
\caption{Spectral image obtained on June 4, 1999. The image has been processed to remove sky and telescope
background, dark current, bad pixels, and pixel-to-pixel variations in response. North is down.
Table~2 gives the approximate wavelength coverage of each complete order, with order 33 being
near the top of the image and order 38 being near the bottom of the image. For reference, note
the strong Br-$\gamma$ (2.166~\micron) emission from ionized gas located near the center of the
image. The CO bandhead begins in the first complete order from the top, about 1/3 of the size
of the image from the left. 
This figure can be found in the full version of the paper at http://www-int.stsci.edu/$\sim$figer/papers/nirspec/vel/ms.ps.
\label{fig-04jus0119_red.fits}}
\end{figure}

\clearpage

\begin{figure}
\epsscale{1.3}
\hspace{3.75in}
\hspace*{1.5in}
\vskip .0in
\caption{SCAM image of the slit-scan region with numerical identifiers \citep{gen00} for each star in Table~3.
Note the shadow of the slit running approximately north-south. 
This figure can be found in the full version of the paper at http://www-int.stsci.edu/$\sim$figer/papers/nirspec/vel/ms.ps.
\label{fig-scam_ids}}
\end{figure}

\clearpage

\begin{figure}
\epsscale{0.9}
\hspace{3.75in}
\hspace*{4.5in}
\vskip .2in
\caption{Rectified order from data displayed in Figure~\ref{fig-04jus0119_red.fits}. The data
have been flipped so that North is up, and labels identify objects in Table~4. Note that the
hot stars (without evidence of CO bandhead absorption) are not identified in the table. 
This figure can be found in the full version of the paper at http://www-int.stsci.edu/$\sim$figer/papers/nirspec/vel/ms.ps.
\label{fig-gc119_rect.fits}}
\end{figure}

\clearpage

\begin{figure}
\epsscale{0.5}
\hspace{3.75in}
\hspace*{4.5in}
\vskip .2in
\caption{Full spectrum of IRS7 over all echelle orders. 
This figure can be found in the full version of the paper at http://www-int.stsci.edu/$\sim$figer/papers/nirspec/vel/ms.ps.
\label{fig-irs7full}}
\end{figure}

\clearpage

\begin{figure}
\epsscale{0.5}
\hspace{3.75in}
\hspace*{4.5in}
\vskip .2in
\caption{Spectrum of IRS7 near the CO bandhead feature, as observed ({\it thin}), and shifted ({\it dotted})
by the velocity corresponding to the result from the cross-correlation with respect to
the spectrum of Arcturus ({\it thick}). The spectra are shifted by arbitrary amounts along
the vertical axis for presentation purposes. This figer can be found in the full version of the paper at http://www-int.stsci.edu/$\sim$figer/papers/nirspec/vel/ms.ps. \label{fig-bothirs7xcor}}
\end{figure}

\clearpage

\begin{figure}
\epsscale{0.5}
\hspace{3.75in}
\hspace*{4.5in}
\vskip .2in
\caption{Spectrum of IRS9 near the CO bandhead feature, as observed ({\it thin}), and shifted ({\it dotted})
by the velocity corresponding to the result from the cross-correlation with respect to
the spectrum of Arcturus ({\it thick}). The spectra are shifted by arbitrary amounts along
the vertical axis for presentation purposes. This figer can be found in the full version of the paper at http://www-int.stsci.edu/$\sim$figer/papers/nirspec/vel/ms.ps. \label{fig-IRS9_gc1high}}
\end{figure}

\clearpage

\begin{figure}
\epsscale{0.5}
\hspace{3.75in}
\hspace*{4.5in}
\vskip .2in
\caption{Spectrum of IRS33E near the CO bandhead feature, as observed ({\it thin}), and shifted ({\it dotted})
by the velocity corresponding to the result from the cross-correlation with respect to
the spectrum of Arcturus ({\it thick}). The spectra are shifted by arbitrary amounts along
the vertical axis for presentation purposes. This figer can be found in the full version of the paper at http://www-int.stsci.edu/$\sim$figer/papers/nirspec/vel/ms.ps. \label{fig-IRS33E_gc1high}}
\end{figure}

\clearpage

\begin{figure}
\epsscale{0.5}
\hspace{3.75in}
\hspace*{4.5in}
\vskip .2in
\caption{Spectrum of IRS20 near the CO bandhead feature, as observed ({\it thin}), and shifted ({\it dotted})
by the velocity corresponding to the result from the cross-correlation with respect to
the spectrum of Arcturus ({\it thick}). The spectra are shifted by arbitrary amounts along
the vertical axis for presentation purposes. This figer can be found in the full version of the paper at http://www-int.stsci.edu/$\sim$figer/papers/nirspec/vel/ms.ps. \label{fig-IRS20_gc1high}}
\end{figure}

\clearpage

\begin{figure}
\epsscale{0.5}
\hspace{3.75in}
\hspace*{4.5in}
\vskip .2in
\caption{Spectrum of IRS12N near the CO bandhead feature, as observed ({\it thin}), and shifted ({\it dotted})
by the velocity corresponding to the result from the cross-correlation with respect to
the spectrum of Arcturus ({\it thick}). The spectra are shifted by arbitrary amounts along
the vertical axis for presentation purposes. This figer can be found in the full version of the paper at http://www-int.stsci.edu/$\sim$figer/papers/nirspec/vel/ms.ps. \label{fig-IRS12N_gc1high}}
\end{figure}

\clearpage

\begin{figure}
\epsscale{1.1}
\hspace{3.75in}
\hspace*{4.5in}
\vskip .2in
\caption{({\it a}) Relationship between the velocity correction for spectral type versus
temperature for template stars.
({\it b}) Relationship between temperature and CO index for template stars. This figer can be found in the full version of the paper at http://www-int.stsci.edu/$\sim$figer/papers/nirspec/vel/ms.ps. \label{fig-vcorr_vs_temperature}}
\end{figure}

\clearpage

\begin{figure}
\epsscale{1.1}
\hspace{3.75in}
\hspace*{4.5in}
\vskip .2in
\caption{({\it a}) Histogram of velocities for cool stars in the central parsec from Table~4.
({\it b}) Same, using data from \citet{gen00}. This figer can be found in the full version of the paper at http://www-int.stsci.edu/$\sim$figer/papers/nirspec/vel/ms.ps. \label{fig-gchist}}
\end{figure}

\clearpage

\begin{figure}
\epsscale{1}
\hspace{5.75in}
\hspace*{4.5in}
\vskip .2in
\caption{Plot showing radial velocities for cool stars in the central parsec from Table~4. Open (filled) circles
correspond to stars with motion away from (toward) the Sun. The area of each circle scales with velocity.
The origin corresponds to the position of Sgr~A*. This figer can be found in the full version of the paper at http://www-int.stsci.edu/$\sim$figer/papers/nirspec/vel/ms.ps. \label{fig-gcbub}}
\end{figure}

\clearpage

\begin{figure}
\epsscale{1.3}
\hspace{4.75in}
\vskip .1in
\caption{Comparison of heliocentric velocities for the 49 stars in common between our sample and that
in \citet{gen00}. Errors for the Genzel et al. data
are shown with error bars. Errors for the data in Table~4 are smaller than the
symbol size. A line with a slope of unity is overplotted. This figer can be found in the full version of the paper at http://www-int.stsci.edu/$\sim$figer/papers/nirspec/vel/ms.ps. \label{fig-gcvsgen}}
\end{figure}

\clearpage

\begin{figure}
\epsscale{1}
\hspace{3.75in}
\hspace*{4.5in}
\vskip .2in
\caption{({\it a}) Radial velocity of stars in the sample as a function of galactic longitude
and ({\it b}) galactic latitude. The horizontal axes are projected distances
from Sgr~A* in parsecs. This figer can be found in the full version of the paper at http://www-int.stsci.edu/$\sim$figer/papers/nirspec/vel/ms.ps. \label{fig-lv}}
\end{figure}

\clearpage

\begin{figure}
\epsscale{1}
\hspace{3.75in}
\hspace*{4.5in}
\vskip .2in
\caption{({\it a})
The thick solid line represents the normalized asymmetry statistic, $\Delta$/N, versus position angle, from
0$\arcdeg$ to 180$\arcdeg$ East of North, as calculated using equation \ref{equation-delta}. $\Delta$ is the number of
stars with negative velocities minus those with positive velocities on one side of the axis
minus the same quantity for stars on the other side of the axis, and N is the total number of stars
in the data set. The dotted and dashed lines
show the number of stars with negative velocities minus those with positive velocities on each
of the two sides of the axis, divided by N. The plots with the highest values show the total number of stars
on each side of the axis. ({\it b}) Cumulative percentage of systems versus maximum normalized $\Delta$ for 1000 simulated
systems and all position angles. Each system has 85 stars with uniform random location within a 1~{\it pc} square area centered on
Sgr~A* and a gaussian velocity distribution having mean of zero and standard deviation of 102~\kms. Note
that 20\% of systems have $\Delta>$0.2. This figer can be found in the full version of the paper at http://www-int.stsci.edu/$\sim$figer/papers/nirspec/vel/ms.ps. \label{fig-gcrot_ours}}
\end{figure}

\begin{figure}
\epsscale{1}
\hspace{3.75in}
\hspace*{4.5in}
\vskip .2in
\caption{
Same as Figure~\ref{fig-gcrot_ours}, except data from \citet{gen00} have been used. The Monte Carlo
simulation uses systems with 198 points, the number of late-type stars in the \citet{gen00}
data set. This figer can be found in the full version of the paper at http://www-int.stsci.edu/$\sim$figer/papers/nirspec/vel/ms.ps. \label{fig-gcrot_genzel}}
\end{figure}

\clearpage

\begin{figure}
\epsscale{1}
\hspace{3.75in}
\hspace*{4.5in}
\vskip .2in
\caption{({\it a}) Velocity dispersion in 5$\arcsec$ wide annuli as a function of radius. ({\it b})
Same, using data from \citet{gen00}. This figer can be found in the full version of the paper at http://www-int.stsci.edu/$\sim$figer/papers/nirspec/vel/ms.ps. \label{fig-vdisp}}
\end{figure}

\clearpage

\begin{figure}
\epsscale{1}
\hspace{3.75in}
\hspace*{4.5in}
\vskip .2in
\caption{({\it a}) Enclosed mass according to equation \ref{equation-BT}
\citep{bah81} using velocities adjusted to the Local Standard of Rest by adding +10~\kms,
as a function of radius. ({\it b}) Same, using data from \citet{gen00},
after subtracting the median velocity of the sample in order
to adjust the data to the LSR. IRS9 has been excluded from the analysis
for both data sets because of its anamolously high velocity. This figer can be found in the full version of the paper at http://www-int.stsci.edu/$\sim$figer/papers/nirspec/vel/ms.ps. \label{fig-mass}}
\end{figure}

\clearpage

\begin{figure}
\epsscale{1.}
\hspace*{4.5in}
\vskip .2in
\caption{({\it a}) Surface number density in our data.
 ({\it b}) Surface number density in \citet{gen00} data.
Neither plot has been adjusted for incompleteness.
\label{fig-radial}}
\end{figure}


\begin{figure}
\epsscale{1.}
\hspace*{4.5in}
\vskip .2in
\caption{({\it a}) Observed and model velocity dispersion in the central parsec.
The model assumes circular velocities, a central dark mass of 3.45(10$^6$)~\Msun,
particles with non-zero mass, and a stellar density distribution
according to equation~\ref{equation-density} with r$_0$=0.4~\pc\ and n=$-$0.25 ({\it solid}) and n=$-$2 ({\it dashed}).
({\it b}) Same, using data in \citet{gen00}.
\label{fig-vdisp_model_n=-0.25_ours_r0=0.4}}
\end{figure}

\begin{figure}
\epsscale{1.}
\hspace*{4.5in}
\vskip .2in
\caption{({\it a}) $\chi^2$ for model compared to observations of the velocity
dispersion as a function of power-law index for stellar density according
to equation~\ref{equation-density} with r$_0$=0.4~\pc.
({\it b}) Same, using data in \citet{gen00}.
\label{fig-chisq_ours_r0=0.4}}
\end{figure}

\begin{figure}
\epsscale{1.}
\hspace*{4.5in}
\vskip .2in
\caption{({\it a}) Locations of 10,000 particles in simulation for 1/r$^2$ distribution.
({\it b}) Same, assuming a distribution
according to equation~\ref{equation-density} with r$_0$=0.4~\pc\ and n=$-$0.25. 
This figure can be found in the full version of the paper at http://www-int.stsci.edu/$\sim$figer/papers/nirspec/vel/ms.ps.
\label{fig-model_0.4circ.xyproj}}
\end{figure}

\clearpage

\begin{figure}
\epsscale{1.}
\hspace*{4.5in}
\vskip .2in
\caption{({\it a}) Projected number density ({\it left axis}) and velocity dispersion ({\it right axis})
of model system, assuming a 1/r$^2$ distribution at all radii.
({\it b}) Same, assuming a distribution
according to equation~\ref{equation-density} with r$_0$=0.4~\pc\ and n=$-$0.25.  
This figure can be found in the full version of the paper at http://www-int.stsci.edu/$\sim$figer/papers/nirspec/vel/ms.ps.
\label{fig-model_0.0circ}}
\end{figure}

\clearpage


\clearpage

\begin{figure}
\epsscale{1}
\hspace{3.75in}
\hspace*{4.5in}
\vskip .2in
\caption{({\it a}) Enclosed mass in the Monte Carlo simulation and that inferred by using
equation \ref{equation-BT} \citep{bah81}, assuming circular orbits,
as a function of radius, and a system having a 1/r$^2$ distribution.
({\it b}) Same, assuming a distribution
according to equation~\ref{equation-density} with r$_0$=0.4~\pc\ and n=$-$0.25. This figer can be found in the full version of the paper at http://www-int.stsci.edu/$\sim$figer/papers/nirspec/vel/ms.ps. \label{fig-model_0.4circ.mass}}
\end{figure}

\clearpage

\begin{figure}
\epsscale{1}
\hspace{3.75in}
\hspace*{4.5in}
\vskip .2in
\caption{Line-of-sight acceleration versus projected distance from Sgr~A* for a star on a circular orbit.
We have assumed a central point mass of 3.45(10$^6$)~\Msun, an orbital  phase of $\pi$/4,
and a range of inclinations between the orbital axis and the line of sight. 
This figure can be found in the full version of the paper at http://www-int.stsci.edu/$\sim$figer/papers/nirspec/vel/ms.ps.
\label{fig-accel}}
\end{figure}

\end{document}